\documentclass[final,3p,twocolumn]{elsarticle}

\usepackage{graphicx}

\biboptions{comma,square,numbers,sort&compress}

\journal{Physics Letters B}

\begin{document}

\begin{frontmatter}

\title{ The Golden Strip of Correlated\\
Top Quark, Gaugino, and Vector-like Mass\\
In No-Scale, No-Parameter ${\cal F}$-$SU(5)$ }

\author[loc_1,loc_2]{Tianjun Li}
\ead{junlt@physics.tamu.edu}

\author[loc_1]{James A. Maxin}
\ead{jmaxin@physics.tamu.edu}

\author[loc_1,loc_3,loc_4]{Dimitri V. Nanopoulos}
\ead{dimitri@physics.tamu.edu}

\author[loc_5]{Joel W. Walker\corref{cor1}}
\ead{jwalker@shsu.edu}
\cortext[cor1]{{\it Corresponding author:} Telephone~+1~(936)~294-4803; Fax~+1~(936)~294-1585}

\address[loc_1]{George P. and Cynthia W. Mitchell Institute for Fundamental Physics,\\
Texas A$\&$M University, College Station, TX 77843, USA }

\address[loc_2]{Key Laboratory of Frontiers in Theoretical Physics, Institute of Theoretical Physics,\\
Chinese Academy of Sciences, Beijing 100190, P. R. China }

\address[loc_3]{Astroparticle Physics Group, Houston Advanced Research Center (HARC),\\
Mitchell Campus, Woodlands, TX 77381, USA}

\address[loc_4]{Academy of Athens, Division of Natural Sciences,\\
28 Panepistimiou Avenue, Athens 10679, Greece }

\address[loc_5]{Department of Physics, Sam Houston State University,
Huntsville, TX 77341, USA }

\begin{abstract}

We systematically establish the hyper-surface
within the $\tan \beta$, top quark mass $m_{\rm t}$, universal gaugino mass $M_{1/2}$,
and vector-like mass $M_{\rm V}$ parameter volume which is compatible with the application
of the No-Scale Supergravity boundary conditions, particularly the vanishing of the Higgs
bilinear soft term $B_\mu$, near to the Planck mass at the point $M_{\cal{F}}$ of ultimate
$\cal{F}$-lipped $SU(5)$ unification. $M_{\cal{F}}$ is elevated from the penultimate partial
unification near the traditional GUT scale at a mass $M_{32}$ by the inclusion of extra
$\cal{F}$-theory derived heavy vector-like multiplets. We demonstrate that simultaneous adherence
to all current experimental constraints, most importantly contributions to the muon
anomalous magnetic moment $(g-2)_\mu$, the branching ratio limit on $(b \rightarrow s\gamma)$,
and the 7-year WMAP relic density measurement, dramatically reduces the allowed solutions
to a highly non-trivial ``golden strip'' with $\tan\beta\simeq 15$,
$m_{\rm t} = 173.0$-$174.4$~GeV, $M_{1/2} = 455$-$481$~GeV, and $M_{\rm V} = 691$-$1020$~GeV,
effectively eliminating all extraneously tunable model parameters.
We emphasize that the consonance of the theoretically viable $m_{\rm t}$ range
with the experimentally established value is an independently correlated ``postdiction''.
The predicted range of $M_{\rm V}$ is testable at the Large Hadron Collider (LHC).
The partial lifetime for proton decay in the leading ${(e\vert\mu)}^{\!+}\! \pi^0$
channels falls around $4.6 \times 10^{34}$~Y,
testable at the future DUSEL and Hyper-Kamiokande facilities.

\end{abstract}


\end{frontmatter}

\section{Introduction}

We have recently demonstrated~\cite{Li:2010ws} the existence of a model
dubbed No-Scale $\cal{F}$-$SU(5)$, resting essentially and in equal
measure on the tripodal foundations of the Flipped $SU(5)\times U(1)_{\rm X}$
Grand Unified Theory (GUT)~\cite{Barr:1981qv,Derendinger:1983aj,Antoniadis:1987dx},
extra $\cal{F}$-theory derived TeV scale vector-like
multiplets~\cite{Jiang:2006hf, Jiang:2009zza, Jiang:2009za, Li:2010dp, Li:2010rz},
and the high scale boundary conditions
of No-Scale Supergravity~\cite{Cremmer:1983bf,Ellis:1983sf,Ellis:1983ei,Ellis:1984bm,Lahanas:1986uc},
which simultaneously satisfies all current experimental
constraints, while eliminating all extraneously tunable free parameters.
The hybridization of these three distinct conceptual progenitors
was shown to uniquely define a ``golden point'' of phenomenological intersection,
at which is {\it dynamically} established the universal gaugino boundary mass $M_{1/2}$,
the ratio of Higgs vacuum expectation values $\tan \beta$, the dual $\cal{F}$-lipped
unification scales $M_{32}$ and $M_{\cal{F}}$, and also consequently the electroweak
symmetry breaking (EWSB) scale, the full contingent of supersymmetric particle masses,
the proton lifetime, and all interrelated experimental observables.

This presentation however, bore the caveat that the mass of the vector-like particles
$M_{\rm V}$ was taken to be a constituent definition of the $\cal{F}$-theory context.
While values around 1~TeV are considered reasonable a priori selections
for natural proximity to the point of EWSB,
consistent decoupling of the traditional GUT and Planck scales,
and even potential testability at the LHC, it was not obvious that similar
constructions with different masses - it has sometimes been suggested that
$M_{\rm V}$ could be as large as $10^{15}$~GeV - might not also exist, thereby smearing
the ``golden point'' into a ``golden string''. We have therefore undertaken
a comprehensive scan of the viable range of this input. 

Concurrently, we set out to establish the effect which variations
within the quoted error margins of the key electroweak (EW) reference data
($\alpha_{\rm s}$,$M_{\rm Z}$) and $m_{\rm t}$ would have on implementation of
the No-Scale $\cal{F}$-$SU(5)$ scenario.
The induced change in $\vert B_\mu(M_{\cal{F}}) \vert$ with respect to shifts in
$\alpha_{\rm s}$ and $M_{\rm Z}$ were mild, on the order of 1 GeV, which
we thus adopted as our definition of acceptable deviation from the strict
$B_{\mu}(M_{\cal{F}}) = 0$ condition, this about the size of the EW
radiative corrections. The variation with respect to $m_{\rm t}$ was found to be
more severe by an order of magnitude, and this we opted instead to recognize 
by effectively treating $m_{\rm t}$ as an additional input, scanning
also over discrete values for this parameter, and selecting
the appropriate $m_{\rm t}$ to restore
compliance with $\vert B_\mu(M_{\cal{F}}) \vert \leq 1$
at each point in the ($\tan \beta$,$M_{1/2}$,$M_{\rm V}$) volume.

As might be expected, we found that the four degrees of freedom
may conspire by intra-compensatory variation to define a large hyper-surface
of acceptable solutions for the No-Scale boundary condition. However, simultaneous
compatibility with experimental results for the branching ratio of $(b\rightarrow s\gamma)$,
the non Standard Model (SM) contribution to $(g-2)_\mu$, and the WMAP 
cold dark matter (CDM) relic density measurement
constitutes a much more stringent condition. The mutually consistent
intersection is a non-trivial ``golden strip'' --
$\tan \beta \simeq 15$, $M_{1/2} = 455$-$481$~GeV, and $M_{\rm V}=691$-$1020$~GeV --
narrowly encompassing our original golden point.

We find the emergent restriction of $M_{\rm V}$
to just a rather light range which may be probed by the LHC
to be quite noteworthy.
The tightly bound GUT couplings and scale
imply a well resolved dimension six partial lifetime
for proton decay in the leading ${(e\vert\mu)}^{\!+}\! \pi^0$
channels around $4.6 \times 10^{34}$~Y, within reach of the future
Hyper-Kamiokande~\cite{Nakamura:2003hk} and Deep Underground Science
and Engineering Laboratory (DUSEL)~\cite{Raby:2008pd} experiments.
A postiori, we recognize the fortune which smiled on our early efforts,
wherein by some chance we may have struck gold on
a first swing of the pick, not realizing that substantial deviation from
the region of $M_{\rm V} \simeq 1$~TeV would destroy the model.

Our greatest surprise and delight however, is reserved for the new findings
regarding $m_{\rm t}$. It seems that under no circumstance is a satisfactory
realization of the present scenario possible unless $m_{\rm t}$, considered
again here as independent input, lies within the range $173.0$-$174.4$~GeV.
The top mass being now well known~\cite{:2009ec}, we firmly resist the temptation to refer to
this result as a ``prediction'', although we consider the pressing impact of the
raw correlation between theoretical and experimental numbers to be undiminished.
The remarkable sonority of this ``postdiction'', which appears only after exhaustion
to the last of all freely tunable model parameters, suggests to us that deeper
currents may be in motion below the surface of No-Scale $\cal{F}$-$SU(5)$.

\section{$\cal{F}$-SU(5) No-Scale Models}

Gauge coupling unification strongly suggests the existence of a GUT.
In minimal supersymmetric $SU(5)$ models 
there are problems with doublet-triplet splitting and dimension
five proton decay by colored Higgsino exchange. These difficulties
can be elegantly overcome in Flipped $SU(5)$ GUT
models via the missing partner mechanism~\cite{Antoniadis:1987dx}.
Written in full, the gauge group of Flipped $SU(5)$ is
$SU(5)\times U(1)_{X}$, which can be embedded into $SO(10)$.
A most notable intrinsic feature of the Flipped $SU(5)$ GUT is the presence
of dual unification scales, with the ultimate merger of $SU(5) \times U(1)_{\rm X}$,
at a scale referred to here as $M_{\cal F}$, occurring subsequent in
energy to the penultimate $SU(3)_{\rm c}$ and $SU(2)_{\rm L}$ mixing at $M_{32}$.

No-Scale Supergravity was proposed~\cite{Cremmer:1983bf,Ellis:1983sf,Ellis:1983ei,Ellis:1984bm,Lahanas:1986uc}
to address the cosmological flatness problem. For the simple K\"ahler potential given in~\cite{Li:2010ws},
we automatically obtain the No-Scale boundary condition $M_0=A=B_{\mu}=0$ on the universal boundary scalar mass
and tri/bi-linear soft terms, while $M_{1/2} > 0$ is allowed, and indeed required for supersymmetry (SUSY) breaking. This appealing
reductionist perspective has however, historically been undermined by a basic inconsistency of
the $M_0 = 0$ condition as applied at a GUT scale of order $10^{16}$~GeV with precision phenomenology.

In the more traditional Flipped $SU(5)$ formulations, the scale $M_{\cal F}$ occurs only
slightly above $M_{32}$, larger by a factor of perhaps only two or three~\cite{Ellis:2002vk}.  Our interest
however, is in scenarios where the ratio $M_{\cal F}/M_{32}$ is considerably larger, on the order of $10$ to $100$.
Key motivations for this picture include the desire to address the monopole problem via hybrid inflation,
and the opportunity for realizing true string scale gauge coupling unification in
the free fermionic model building context~\cite{Jiang:2006hf, Lopez:1992kg},
or the decoupling scenario in F-theory models~\cite{Jiang:2009zza,Jiang:2009za}.
We have previously also considered the favorable effect of such considerations
on the decay rate of the proton~\cite{Li:2010dp,Li:2010rz}.

The greatest present benefit however, is the effortless manner in which the lifting of the $SU(5) \times U(1)_{\rm X}$
scale salvages the dynamically established boundary conditions of No-Scale Supergravity.  Being highly predictive, 
these conditions are thus also intrinsically highly constrained, and notoriously difficult to realize generically.
Our continuing study, succinctly dubbed No-Scale ${\cal F}$-$SU(5)$~\cite{Jiang:2006hf, Jiang:2009zza, Jiang:2009za, Li:2010dp, Li:2010rz},
of the ${\cal F}$-lipped $SU(5)$ GUT~\cite{Lopez:1992kg} supplemented by ${\cal F}$-theory derived vector-like multiplets
at the TeV scale, provides the essential rationale; The accompanying modification to the gauge coupling renormalization group
equations (RGEs) naturally separates $M_{32} \simeq 1.0 \times 10^{16}$ GeV, near the traditional GUT scale,
from $M_{\cal F} \simeq 7.5\times 10^{17}$GeV, approaching the reduced Planck mass~\cite{Jiang:2006hf, Jiang:2009zza, Jiang:2009za},
at which point the No-Scale boundary conditions fit like hand to glove.

In this scenario, we introduce the following two pairs 
of vector-like Flipped $SU(5)\times U(1)_X$ multiplets
near the TeV scale~\cite{Jiang:2006hf}
\begin{eqnarray}
\hspace{-.3in}
& \left( {XF}_{\mathbf{(10,1)}} \equiv (XQ,XD^c,XN^c),~{\overline{XF}}_{\mathbf{({\overline{10}},-1)}} \right)\, ,&
\nonumber \\
\hspace{-.3in}
& \left( {Xl}_{\mathbf{(1, -5)}},~{\overline{Xl}}_{\mathbf{(1, 5)}}\equiv XE^c \right)\, ,&
\label{z1z2}
\end{eqnarray}
where $XQ$, $XD^c$, $XE^c$, $XN^c$ have the same quantum numbers as the
quark doublet, the right-handed down-type quark, charged lepton, and
neutrino, respectively. We thus suggest the name $\textit{Flippons}$ for these hypothetical particles.

We emphasize that the specific representations of vector-like fields which we currently employ have been explicitly
constructed within the F-theory model building context~\cite{Jiang:2009zza}.
However, the mass of these fields, and even the fact of their existence, is not mandated by the F-theory, wherein it is also possible to
realize models with only the traditional Flipped (or Standard) $SU(5)$ field content.  We claim only an inherent consistency of their conceptual
origin out of the F-theoretic construction, and take the manifest phenomenological benefits which accompany the elevation of
$M_{\cal F}$ as justification for the greater esteem which we hold for this particular model above other alternatives.
In our present point of view, the uniqueness of the theory is imposed by the experimental constraints, punctuated by
the relative efficiency with which No-Scale ${\cal F}$-$SU(5)$ achieves such agreement, in contrast to its competitors.

There are, however, delicate questions of compatibility between the F-theoretic model building origins
of $\cal{F}$-$SU(5)$ with vector-like fields, and the purely field-theoretic RGE running which
we employ up to the high scale.  As one approaches the Planck scale $M_{\rm Pl}$,
consideration must be given to the role which will be played by Kaluza Klein (KK) and string mode
excitations, and if we indeed posit a substantial increase in the string scale $M_{\rm S}$,
also to $\alpha^\prime$ corrections associated with the corresponding reduction in the
global volume of the six-dimensional internal space via the scaling
$R_{\rm Global} \propto {( M_{\rm Pl}/M_{\rm S} )}^{1/3}$.

For local F-theory models with large volume compactifications, we
acknowledge that the string scale determined by direct calculation cannot be
large~\cite{Blumenhagen:2009gk}.  However, to describe Nature, the local F-theory
models must be embedded into a globally consistent framework.
In such global constructions, the string and KK mass scales can indeed be
comfortably positioned around $4 \times 10^{17}$~GeV,
as is likewise the case with the usual heterotic string constructions.
It should be remarked in any event, that since the running of the gauge
couplings is logarithmically dependent upon the mass scale, the contributions
to the RGEs from the string and KK mode excitations are quite small.
Moreover, there may further exist contributions to the RGE running of the gauge
couplings from the heavy threshold corrections of heavy fields, as studied for example
in Ref.~\cite{Jiang:2009za}, where the Type IA1 Model of Table X is associated
with a unification about $2 \times 10^{17}$~GeV, somewhat below the usual string scale.

The most important question is whether our model can in fact be embedded into
a globally consistent framework.  It seems to us that a field-theoretic application
of the no-scale boundary conditions may prove to be valid in this case.
This is a point which we continue to study, and on which we have not yet
reached a definitive conclusion.  Therefore, the above potential stringy modifications duly noted,
we stipulate here by choice to consider the minimal possible scenario, wherein their substantive onset
is deferred to $M_{\cal F}$, the naturally elevated secondary unification
point of No-Scale $\cal{F}$-$SU(5)$, and the true GUT scale of this model.
The consequent phenomenology is in our view of sufficient merit and interest to justify
the investigation of the model on its own terms, postponing for now the consideration
of what may ultimately be deemed to constitute effects of higher order.

\section{The Golden Strip}

In the No-Scale formulation, one imposes $M_0 = A = B_\mu$ = 0 at the unification scale
$M_{\cal F}$, and allows distinct inputs for the single parameter $M_{1/2}(M_{\cal F})$ to
translate under RGEs to distinct low scale outputs.
Equivalently, we instead allow $M_{1/2}$ and $\tan \beta$ to float freely
and implement a precision self-consistency assessment~\cite{Li:2010ws},
customized from the codebase of~\cite{Djouadi:2002ze} and~\cite{Belanger:2008sj},
to isolate solutions for $B_\mu(M_{\cal F}) = 0$.
We adhere to the following experimental constraints:
1) WMAP 7-year measurements of 
the CDM density~\cite{Komatsu:2010fb}, 
0.1088 $\leq \Omega_{\chi} \leq$ 0.1158. 
2) Experimental limits on 
the FCNC process, $b \rightarrow s\gamma$, using the limits
$2.86 \times 10^{-4} \leq Br(b \rightarrow s\gamma) 
\leq 4.18 \times 10^{-4}$~\cite{Barberio:2007cr, Misiak:2006zs}.
3) Anomalous magnetic moment of the muon, $g_{\mu} - 2$, with a 
lower bound of 
$a_{\mu} > 11 \times 10^{-10}$~\cite{Bennett:2004pv}.
4) Process $B_{s}^{0} \rightarrow \mu^+ \mu^-$, using an upper bound of 
$Br(B_{s}^{0} \rightarrow \mu^{+}\mu^{-}) < 5.8 \times 10^{-8}$~\cite{:2007kv}. 
5) LEP limit on the lightest CP-even Higgs boson 
mass, $m_{h} \geq 114$ GeV~\cite{Barate:2003sz,Yao:2006px}.

\begin{figure}[ht]
	\centering
	\includegraphics[width=0.50\textwidth]{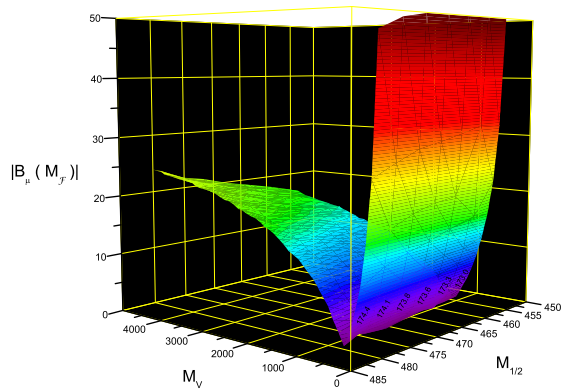}
	\caption{
The $B_\mu = 0$ target for
variations in $(M_{1/2},M_{\rm V})$, with $\tan \beta = 15$.
The specific $m_{\rm t}$ which is required to minimize
$\vert B_\mu(M_{\cal{F}}) \vert$ is annotated along the solution string.
	}
        \label{fig:bvalley}
\end{figure}

Only a small portion of viable
parameter space is consistent with
the $B_{\mu}(M_{\cal F}) = 0$ condition, which thus
constitutes a strong constraint. Since the boundary value of the universal gaugino mass $M_{1/2}$, and
even the unification scale $M_{\cal F} \simeq 7.5 \times 10^{17}$ GeV itself, are
established by the low energy experiments via
RGE running, we are not left with
any surviving scale parameters in the present model. The floor of the ``valley gorge" in
Fig.~{\ref{fig:bvalley}} represents accord with the $B_\mu = 0$ target for
variations in $(M_{1/2},M_{\rm V})$. We fix $\tan \beta = 15$,
as appears to be rather generically required in No-Scale $\cal{F}$-$SU(5)$
to realize radiative EWSB and match the observed CDM density.

\begin{figure*}[ht]
	\centering
	\includegraphics[width=0.51\textwidth]{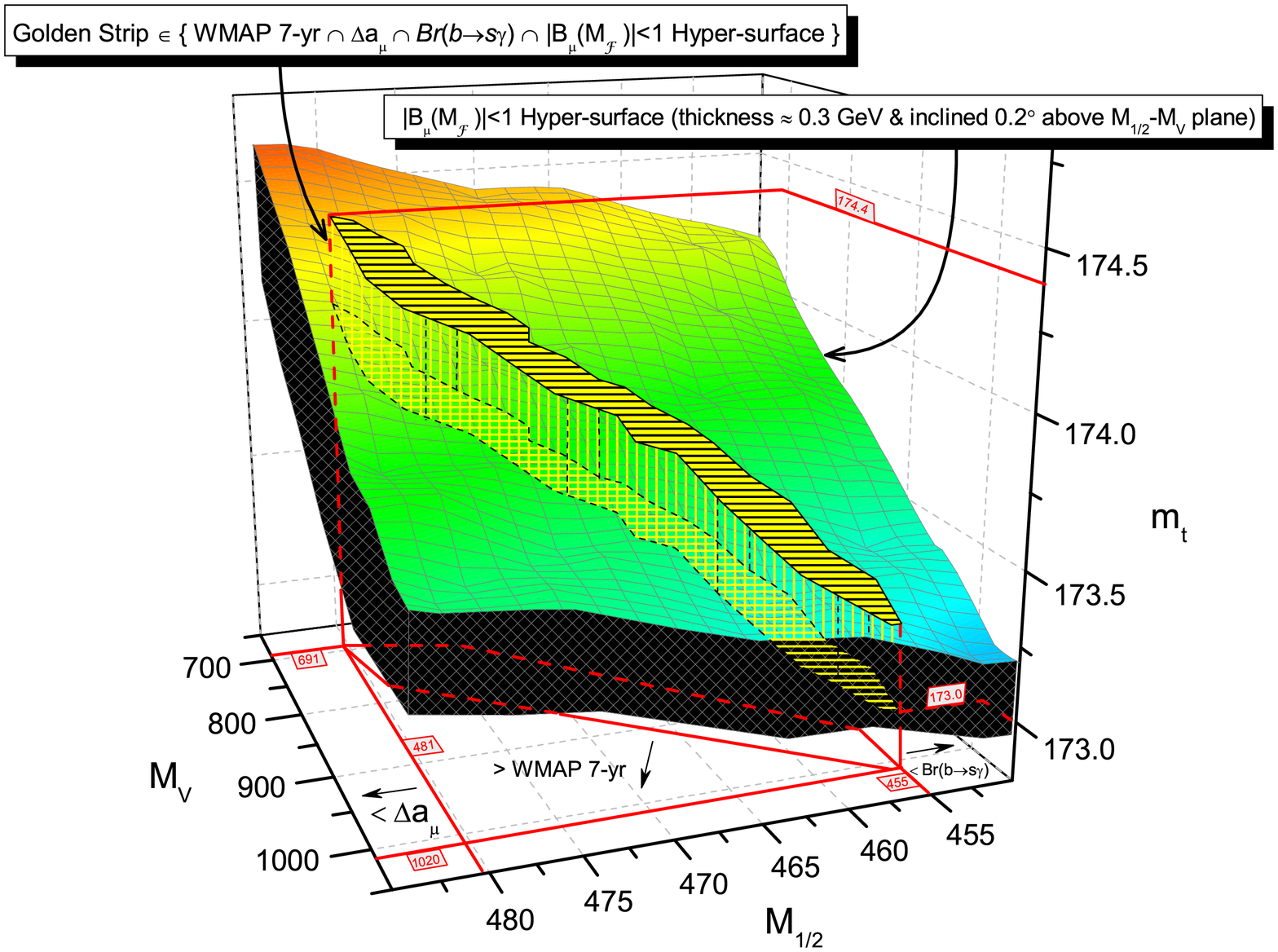}
	\hspace{0.05\textwidth}
	\includegraphics[width=0.39\textwidth]{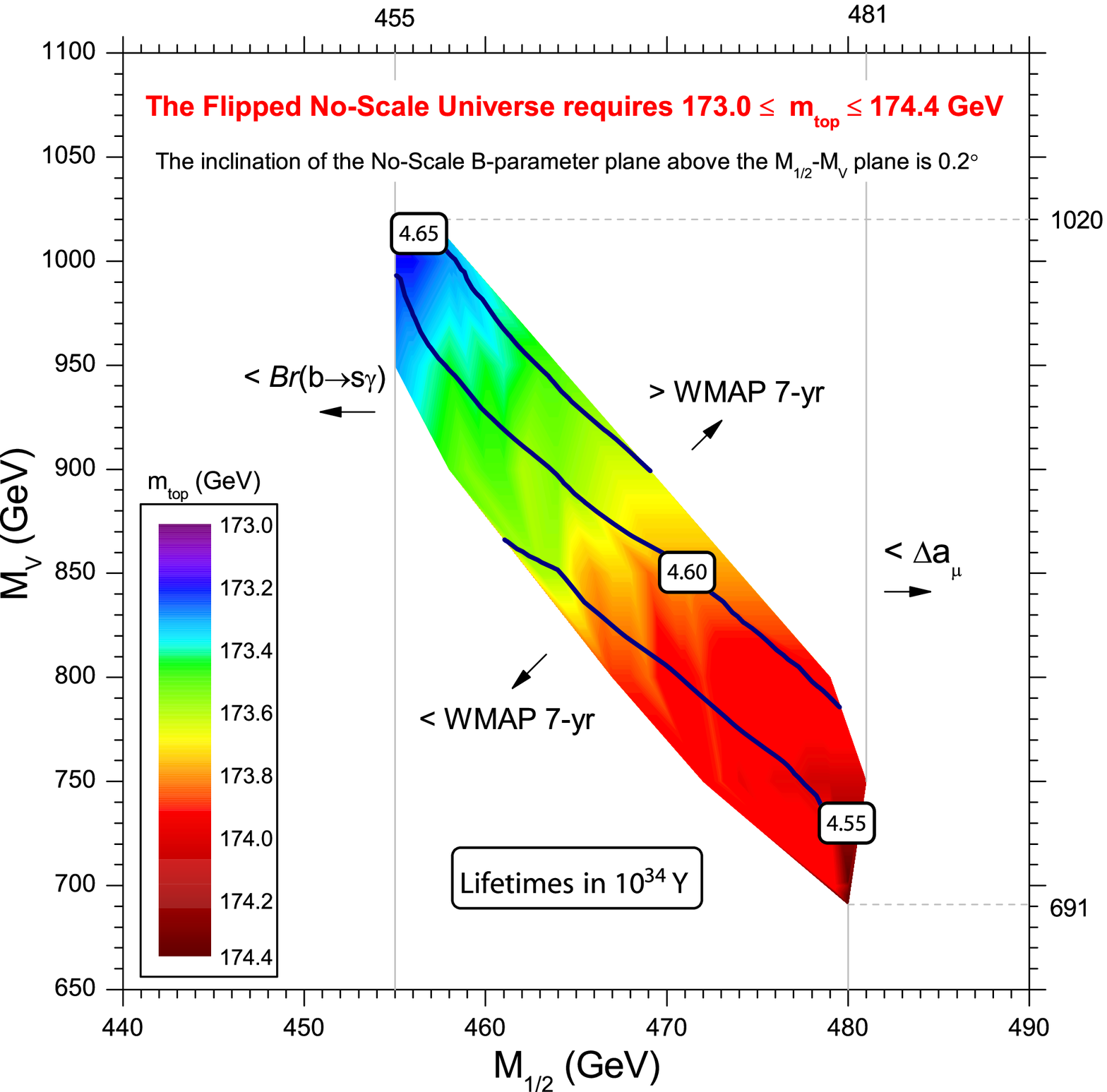}
	\caption{
With $\tan \beta \simeq 15$ fixed by WMAP-7, the residual parameter volume is three dimensional
in $(M_{12},M_{\rm V},m_{\rm t})$, with the $\vert B_\mu(M_{\cal{F}}) \vert \leq 1$ (slightly thickened)
surface forming a shallow $(0.2^\circ)$ incline above the $(M_{12},M_{\rm V})$ plane.
The overlayed blue contours on the flattened diagram
mark the $p \!\rightarrow\! {(e\vert\mu)}^{\!+}\! \pi^0$ proton lifetime
prediction, in units of $10^{34}$ years.
	}
        \label{fig:goldenstrip}
\end{figure*}

We have allowed for uncertainty
in the most sensitive experimental input, the top quark mass,
by effectively redefining $m_{\rm t}$ as an independent free parameter.
Lesser sensitivities to uncertainty in $(\alpha_{\rm s},M_{\rm Z})$ are included
in the $\pm 1$~GeV deviation from strict adherence to $B_\mu = 0$.
We have established that there is a two dimensional sheet (of some marginal thickness
to recognize the mentioned uncertainty) defining $\vert B_\mu(M_{\cal{F}}) \vert \leq 1$
for each point in the three dimensional ($M_{1/2}$, $M_{\rm V}$, $m_{\rm t}$) volume,
as shown in Figs.~(\ref{fig:goldenstrip}).
This sheet is inclined in the region of interest at the very shallow angle of
$0.2^\circ$ to the ($M_{1/2}$,$M_{\rm V}$) plane, such that $m_{\rm t}$ is largely decoupled
from variation in the plane.

A particularly interesting facet of this model is the surprisingly strong
correlation between the vector-like mass $M_{\rm V}$ and the WMAP dark matter relic density.
The underlying mechanism may be traced to the more directly visible effect which $M_{\rm V}$
has on the secondary unification scale $M_{\cal F}$.  A larger (or smaller) value of $M_{\rm V}$
will reduce (or increase) $M_{\cal F}$, thus suppressing (or enhancing) the the role of the
RGEs across the contracted (or extended) energy gap.  A key consequence will be a heavier
(or lighter) low energy Bino mass.  In our model, which features $99.8 \%$ Bino dark matter,
the dark matter density, being proportional to the Bino mass squared, will thus also sharply
rise (or fall).  For a fixed $M_{1/2}$, we may thus place upper and lower bounds on $M_{\rm V}$
via the corresponding upper and lower limits of the WMAP dark matter density.

The $(g-2)_\mu$ and $b\rightarrow s\gamma$ constraints
vary most strongly with $M_{1/2}$.
The two considered effects are each at their lower limits at the boundary,
but they exert pressure in opposing directions on $M_{1/2}$ due to the fact that the
leading gaugino and squark contributions to $Br(b\rightarrow s\gamma)$ enter with an opposing sign to the SM
term and Higgs contribution. For the non-SM contribution to $\Delta a_\mu$, the effect is additive,
and establishes an upper mass limit on $M_{1/2}$.
Incidentally, the same experiment forms the central rationale for the adoption of
$\rm{sign}(\mu) > 0$, such that appropriate interference terms between SM and SUSY
contributions are realized.
Conversely, the requirement that SUSY contributions to $Br(b\rightarrow s \gamma)$ not be overly large,
undoing the SM effect, requires a sufficiently large, {\it i.~e.}~lower bounded, $M_{1/2}$.
The WMAP-7 CDM measurement, by contrast, exhibits a fairly strong correlation with both
$M_{1/2}$ and $M_{\rm V})$ (as elaborated prior), cross-cutting the $M_{1/2}$ bound, and confining the vector-like mass
to 691-1020 GeV. We note that the mixing of the SM fermions and vector particles may give
additional contributions to $Br(b\rightarrow s\gamma)$ and $\Delta a_\mu$, but we do not consider them here.

The intersection of these three key constraints with the
$\vert B_\mu(M_{\cal{F}}) \vert \leq 1$ surface, as depicted in
Figs.~(\ref{fig:goldenstrip}), defines the
``golden strip'' of No-Scale $\cal{F}$-$SU(5)$.
All of the prior is accomplished with no reference
to the experimental top quark mass, redefined here as a {\it free} input.
However, the extremely shallow angle of inclination ($0.2^\circ$) of
the $\vert B_\mu(M_{\cal{F}}) \vert \leq 1$ sheet
above the $(M_{1/2},M_{\rm V})$ plane and into the $m_{\rm t}$ axis
causes the golden strip to {\it imply} an exceedingly narrow
range of compatibility for $m_{\rm t}$, between $173.0$-$174.4$~GeV,
in perfect alignment with the physically observed value
of $m_{\rm t}$ = $173.1\pm 1.3$ GeV~\cite{:2009ec}.

Within the golden strip, we select the benchmark 
point of Table~\ref{tab:masses}. The golden strip is further consistent with the CDMSII~\cite{Ahmed:2008eu} and 
Xenon100~\cite{Aprile:2010um}
upper limits, with the spin-independent cross section extending from 
$\sigma_{SI} = 1.3$-$1.9 \times 10^{-10}$ pb.
Likewise, the allowed region satisfies the 
Fermi-LAT space telescope constraints~\cite{Abdo:2010dk}, with the 
photon-photon annihilation cross section 
$\left\langle \sigma v \right\rangle_{\gamma\gamma}$ ranging 
from $\left\langle \sigma v \right\rangle_{\gamma\gamma} = 1.5$-$1.7 \times 10^{-28} ~cm^{3}/s$.

\begin{table*}[htb]
	\centering
	\caption{Spectrum (in GeV) for the benchmark point. 
Here, $M_{1/2}$ = 464 GeV, $M_{V}$ = 850 GeV, $m_{t}$ = 173.6 GeV, $\Omega_{\chi}$ = 0.112, $\sigma_{SI} = 1.7 \times 10^{-10}$ pb, and
$\left\langle \sigma v \right\rangle_{\gamma\gamma} = 1.7 \times 10^{-28} ~cm^{3}/s$.
The central prediction for the $p \!\rightarrow\! {(e\vert\mu)}^{\!+}\! \pi^0$ 
proton lifetime is $4.6 \times 10^{34}$ years. The lightest neutralino is 99.8\% Bino.}
	\begin{tabular}{|c|c||c|c||c|c||c|c||c|c||c|c|} \hline		
$\widetilde{\chi}_{1}^{0}$&$96$&$\widetilde{\chi}_{1}^{\pm}$&$187$&$\widetilde{e}_{R}$&$153$&$\widetilde{t}_{1}$&$499$&$\widetilde{u}_{R}$&$975$&$m_{h}$&$120.6$\\ \hline
$\widetilde{\chi}_{2}^{0}$&$187$&$\widetilde{\chi}_{2}^{\pm}$&$849$&$\widetilde{e}_{L}$&$519$&$\widetilde{t}_{2}$&$929$&$\widetilde{u}_{L}$&$1062$&$m_{A,H}$&$946$\\ \hline
$\widetilde{\chi}_{3}^{0}$&$845$&$\widetilde{\nu}_{e/\mu}$&$513$&$\widetilde{\tau}_{1}$&$105$&$\widetilde{b}_{1}$&$880$&$\widetilde{d}_{R}$&$1018$&$m_{H^{\pm}}$&$948$\\ \hline
$\widetilde{\chi}_{4}^{0}$&$848$&$\widetilde{\nu}_{\tau}$&$506$&$\widetilde{\tau}_{2}$&$514$&$\widetilde{b}_{2}$&$992$&$\widetilde{d}_{L}$&$1065$&$\widetilde{g}$&$629$\\ \hline
	\end{tabular}
	\label{tab:masses}
\end{table*}

\section{Experimental Signature}

We remark in closing on a distinctive evidentiary ``smoking gun'' for the No-Scale
$\cal{F}$-$SU(5)$ scenario. 
This is direct detection near the TeV scale
of the components of the extra vector-like flippon multiplets of Eq.~(1). In particular,
these vector particles mirror the quark and lepton quantum numbers,
and crucially also the ``flipped'' charge assignment. Since the
flippons consist of a pair of ten-plets ($XF,\overline{XF}$),
and a pair of {\it charged} $SU(5)$ singlets ($Xl,\overline{Xl}$),
but no five-plets, the grouping is unambiguous. The discovery of flippons would firmly establish the flipped group structure.

\section{Discussion and Conclusion}

We have continued and extended our prior study~\cite{Li:2010ws} of No-Scale Supergravity
in the context of a ${\cal F}$-lipped $SU(5)\times U(1)_X$ GUT
supplemented with ${\cal F}$-theory derived TeV-scale vector-like particles.
There did not have to be an experimentally viable $B_\mu(M_{\cal F}) = 0$
solution at all, and indeed successful implementation of this boundary has eluded a
myriad of prior attempts. Because the universal gaugino mass, and even the
final unification scale $M_{\cal{F}}$ itself are determined
by the low energy known experiments via self consistent RGE running,
there are no surviving arbitrary mass scales or extraneously tunable inputs.
We stress again that the union of top-down
model based constraints with bottom-up experimental data exhausts
the available freedom of parameterization in a uniquely consistent and predictive manner,
{\it prior} to invocation of the $m_{\rm t}$ value. Retaining no residual malleability,
the model is forced to live or die by the success of its extraordinarily finely
attenuated postdiction of the top quark mass -- a trial which it surmounts
with colors flying, phenomenologically defining a ``golden strip'' of correlated
top quark, gaugino, and vector-like mass, with $m_{\rm t} = 173.0$-$174.4$~GeV,
$M_{1/2} = 455$-$481$~GeV, and $M_{\rm V} = 691$-$1020$~GeV. A narrowly defined
yet broadly applicable prediction has been made for $\tan\beta\simeq 15$.
The required TeV scale vector multiplets and dimension six ${(e\vert\mu)}^{\!+}\! \pi^0$
proton decay, both bearing the distinctive signature of their flipped origin,
are each poised to play a potentially prominent role
in certain of the most exciting particle physics experiments of the coming decade.
This luxury of portent and paucity of accommodation 
is the power of No-Scale ${\cal F}$-$SU(5)$.

\section{Acknowledgments}

This research was supported in part 
by the DOE grant DE-FG03-95-Er-40917 (TL and DVN),
by the Natural Science Foundation of China 
under grant No. 10821504 (TL),
and by the Mitchell-Heep Chair in High Energy Physics (TL).

\bibliographystyle{model1-num-names}

\bibliography{bibliography.bib}

\begin{thebibliography}{32}
\expandafter\ifx\csname natexlab\endcsname\relax\def\natexlab#1{#1}\fi
\providecommand{\bibinfo}[2]{#2}
\ifx\xfnm\relax \def\xfnm[#1]{\unskip,\space#1}\fi
\bibitem[{Li et~al.(2010)Li, Maxin, Nanopoulos, and Walker}]{Li:2010ws}
\bibinfo{author}{T.~Li}, \bibinfo{author}{J.~A. Maxin}, \bibinfo{author}{D.~V.
  Nanopoulos}, \bibinfo{author}{J.~W. Walker},
\newblock \bibinfo{title}{{The Golden Point of No-Scale and No-Parameter ${\cal
  F}$-$SU(5)$}}  (\bibinfo{year}{2010}).
\bibitem[{Barr(1982)}]{Barr:1981qv}
\bibinfo{author}{S.~M. Barr},
\newblock \bibinfo{title}{{A New Symmetry Breaking Pattern for $SO(10)$ and
  Proton Decay}},
\newblock \bibinfo{journal}{Phys. Lett.} \bibinfo{volume}{B112}
  (\bibinfo{year}{1982}) \bibinfo{pages}{219}.
\bibitem[{Derendinger et~al.(1984)Derendinger, Kim, and
  Nanopoulos}]{Derendinger:1983aj}
\bibinfo{author}{J.~P. Derendinger}, \bibinfo{author}{J.~E. Kim},
  \bibinfo{author}{D.~V. Nanopoulos},
\newblock \bibinfo{title}{{Anti-$SU(5)$}},
\newblock \bibinfo{journal}{Phys. Lett.} \bibinfo{volume}{B139}
  (\bibinfo{year}{1984}) \bibinfo{pages}{170}.
\bibitem[{Antoniadis et~al.(1987)Antoniadis, Ellis, Hagelin, and
  Nanopoulos}]{Antoniadis:1987dx}
\bibinfo{author}{I.~Antoniadis}, \bibinfo{author}{J.~R. Ellis},
  \bibinfo{author}{J.~S. Hagelin}, \bibinfo{author}{D.~V. Nanopoulos},
\newblock \bibinfo{title}{{Supersymmetric Flipped $SU(5)$ Revitalized}},
\newblock \bibinfo{journal}{Phys. Lett.} \bibinfo{volume}{B194}
  (\bibinfo{year}{1987}) \bibinfo{pages}{231}.
\bibitem[{Jiang et~al.(2007)Jiang, Li, and Nanopoulos}]{Jiang:2006hf}
\bibinfo{author}{J.~Jiang}, \bibinfo{author}{T.~Li}, \bibinfo{author}{D.~V.
  Nanopoulos},
\newblock \bibinfo{title}{{Testable Flipped $SU(5) \times U(1)_X$ Models}},
\newblock \bibinfo{journal}{Nucl. Phys.} \bibinfo{volume}{B772}
  (\bibinfo{year}{2007}) \bibinfo{pages}{49--66}.
\bibitem[{Jiang et~al.(2009)Jiang, Li, Nanopoulos, and Xie}]{Jiang:2009zza}
\bibinfo{author}{J.~Jiang}, \bibinfo{author}{T.~Li}, \bibinfo{author}{D.~V.
  Nanopoulos}, \bibinfo{author}{D.~Xie},
\newblock \bibinfo{title}{{F-$SU(5)$}},
\newblock \bibinfo{journal}{Phys. Lett.} \bibinfo{volume}{B677}
  (\bibinfo{year}{2009}) \bibinfo{pages}{322--325}.
\bibitem[{Jiang et~al.(2010)Jiang, Li, Nanopoulos, and Xie}]{Jiang:2009za}
\bibinfo{author}{J.~Jiang}, \bibinfo{author}{T.~Li}, \bibinfo{author}{D.~V.
  Nanopoulos}, \bibinfo{author}{D.~Xie},
\newblock \bibinfo{title}{{Flipped $SU(5) \times U(1)_X$ Models from
  F-Theory}},
\newblock \bibinfo{journal}{Nucl. Phys.} \bibinfo{volume}{B830}
  (\bibinfo{year}{2010}) \bibinfo{pages}{195--220}.
\bibitem[{Li et~al.(2011)Li, Nanopoulos, and Walker}]{Li:2010dp}
\bibinfo{author}{T.~Li}, \bibinfo{author}{D.~V. Nanopoulos},
  \bibinfo{author}{J.~W. Walker},
\newblock \bibinfo{title}{{Elements of F-ast Proton Decay}},
\newblock \bibinfo{journal}{Nucl. Phys.} \bibinfo{volume}{B846}
  (\bibinfo{year}{2011}) \bibinfo{pages}{43--99}.
\bibitem[{Li et~al.(2010)Li, Maxin, Nanopoulos, and Walker}]{Li:2010rz}
\bibinfo{author}{T.~Li}, \bibinfo{author}{J.~A. Maxin}, \bibinfo{author}{D.~V.
  Nanopoulos}, \bibinfo{author}{J.~W. Walker},
\newblock \bibinfo{title}{{Dark Matter, Proton Decay and Other Phenomenological
  Constraints in ${\cal F}$-$SU(5)$}}  (\bibinfo{year}{2010}).
\bibitem[{Cremmer et~al.(1983)Cremmer, Ferrara, Kounnas, and
  Nanopoulos}]{Cremmer:1983bf}
\bibinfo{author}{E.~Cremmer}, \bibinfo{author}{S.~Ferrara},
  \bibinfo{author}{C.~Kounnas}, \bibinfo{author}{D.~V. Nanopoulos},
\newblock \bibinfo{title}{{Naturally Vanishing Cosmological Constant in $N=1$
  Supergravity}},
\newblock \bibinfo{journal}{Phys. Lett.} \bibinfo{volume}{B133}
  (\bibinfo{year}{1983}) \bibinfo{pages}{61}.
\bibitem[{Ellis et~al.(1984{\natexlab{a}})Ellis, Lahanas, Nanopoulos, and
  Tamvakis}]{Ellis:1983sf}
\bibinfo{author}{J.~R. Ellis}, \bibinfo{author}{A.~B. Lahanas},
  \bibinfo{author}{D.~V. Nanopoulos}, \bibinfo{author}{K.~Tamvakis},
\newblock \bibinfo{title}{{No-Scale Supersymmetric Standard Model}},
\newblock \bibinfo{journal}{Phys. Lett.} \bibinfo{volume}{B134}
  (\bibinfo{year}{1984}{\natexlab{a}}) \bibinfo{pages}{429}.
\bibitem[{Ellis et~al.(1984{\natexlab{b}})Ellis, Kounnas, and
  Nanopoulos}]{Ellis:1983ei}
\bibinfo{author}{J.~R. Ellis}, \bibinfo{author}{C.~Kounnas},
  \bibinfo{author}{D.~V. Nanopoulos},
\newblock \bibinfo{title}{{Phenomenological $SU(1,1)$ Supergravity}},
\newblock \bibinfo{journal}{Nucl. Phys.} \bibinfo{volume}{B241}
  (\bibinfo{year}{1984}{\natexlab{b}}) \bibinfo{pages}{406}.
\bibitem[{Ellis et~al.(1984{\natexlab{c}})Ellis, Kounnas, and
  Nanopoulos}]{Ellis:1984bm}
\bibinfo{author}{J.~R. Ellis}, \bibinfo{author}{C.~Kounnas},
  \bibinfo{author}{D.~V. Nanopoulos},
\newblock \bibinfo{title}{{No Scale Supersymmetric Guts}},
\newblock \bibinfo{journal}{Nucl. Phys.} \bibinfo{volume}{B247}
  (\bibinfo{year}{1984}{\natexlab{c}}) \bibinfo{pages}{373--395}.
\bibitem[{Lahanas and Nanopoulos(1987)}]{Lahanas:1986uc}
\bibinfo{author}{A.~B. Lahanas}, \bibinfo{author}{D.~V. Nanopoulos},
\newblock \bibinfo{title}{{The Road to No Scale Supergravity}},
\newblock \bibinfo{journal}{Phys. Rept.} \bibinfo{volume}{145}
  (\bibinfo{year}{1987}) \bibinfo{pages}{1}.
\bibitem[{Nakamura(2003)}]{Nakamura:2003hk}
\bibinfo{author}{K.~Nakamura},
\newblock \bibinfo{title}{{Hyper-Kamiokande: A next generation water Cherenkov
  detector}},
\newblock \bibinfo{journal}{Int. J. Mod. Phys.} \bibinfo{volume}{A18}
  (\bibinfo{year}{2003}) \bibinfo{pages}{4053--4063}.
\bibitem[{Raby et~al.(2008)}]{Raby:2008pd}
\bibinfo{author}{S.~Raby}, et~al.,
\newblock \bibinfo{title}{{DUSEL Theory White Paper}}  (\bibinfo{year}{2008}).
\bibitem[{:20(2009)}]{:2009ec}
\bibinfo{title}{{Combination of CDF and D\O Results on the Mass of the Top
  Quark}}  (\bibinfo{year}{2009}).
\bibitem[{Ellis et~al.(2002)Ellis, Nanopoulos, and Walker}]{Ellis:2002vk}
\bibinfo{author}{J.~R. Ellis}, \bibinfo{author}{D.~V. Nanopoulos},
  \bibinfo{author}{J.~Walker},
\newblock \bibinfo{title}{{Flipping SU(5) out of trouble}},
\newblock \bibinfo{journal}{Phys. Lett.} \bibinfo{volume}{B550}
  (\bibinfo{year}{2002}) \bibinfo{pages}{99--107}.
\bibitem[{Lopez et~al.(1993)Lopez, Nanopoulos, and Yuan}]{Lopez:1992kg}
\bibinfo{author}{J.~L. Lopez}, \bibinfo{author}{D.~V. Nanopoulos},
  \bibinfo{author}{K.-j. Yuan},
\newblock \bibinfo{title}{{The Search for a realistic flipped $SU(5)$ string
  model}},
\newblock \bibinfo{journal}{Nucl. Phys.} \bibinfo{volume}{B399}
  (\bibinfo{year}{1993}) \bibinfo{pages}{654--690}.
\bibitem[{Blumenhagen et~al.(2009)Blumenhagen, Conlon, Krippendorf, Moster, and
  Quevedo}]{Blumenhagen:2009gk}
\bibinfo{author}{R.~Blumenhagen}, \bibinfo{author}{J.~P. Conlon},
  \bibinfo{author}{S.~Krippendorf}, \bibinfo{author}{S.~Moster},
  \bibinfo{author}{F.~Quevedo},
\newblock \bibinfo{title}{{SUSY Breaking in Local String/F-Theory Models}},
\newblock \bibinfo{journal}{JHEP} \bibinfo{volume}{09} (\bibinfo{year}{2009})
  \bibinfo{pages}{007}.
\bibitem[{Djouadi et~al.(2007)Djouadi, Kneur, and Moultaka}]{Djouadi:2002ze}
\bibinfo{author}{A.~Djouadi}, \bibinfo{author}{J.-L. Kneur},
  \bibinfo{author}{G.~Moultaka},
\newblock \bibinfo{title}{{SuSpect: A Fortran code for the supersymmetric and
  Higgs particle spectrum in the MSSM}},
\newblock \bibinfo{journal}{Comput. Phys. Commun.} \bibinfo{volume}{176}
  (\bibinfo{year}{2007}) \bibinfo{pages}{426--455}.
\bibitem[{Belanger et~al.(2009)Belanger, Boudjema, Pukhov, and
  Semenov}]{Belanger:2008sj}
\bibinfo{author}{G.~Belanger}, \bibinfo{author}{F.~Boudjema},
  \bibinfo{author}{A.~Pukhov}, \bibinfo{author}{A.~Semenov},
\newblock \bibinfo{title}{{Dark matter direct detection rate in a generic model
  with micrOMEGAs2.1}},
\newblock \bibinfo{journal}{Comput. Phys. Commun.} \bibinfo{volume}{180}
  (\bibinfo{year}{2009}) \bibinfo{pages}{747--767}.
\bibitem[{Komatsu et~al.(2010)}]{Komatsu:2010fb}
\bibinfo{author}{E.~Komatsu}, et~al.,
\newblock \bibinfo{title}{{Seven-Year Wilkinson Microwave Anisotropy Probe
  (WMAP) Observations: Cosmological Interpretation}}  (\bibinfo{year}{2010}).
\bibitem[{Barberio et~al.(2007)}]{Barberio:2007cr}
\bibinfo{author}{E.~Barberio}, et~al.,
\newblock \bibinfo{title}{{Averages of $b-$hadron properties at the end of
  2006}}  (\bibinfo{year}{2007}).
\bibitem[{Misiak et~al.(2007)}]{Misiak:2006zs}
\bibinfo{author}{M.~Misiak}, et~al.,
\newblock \bibinfo{title}{{The first estimate of Br$(\overline{B} \rightarrow
  X_{s} \gamma)$ at ${\cal O}({\alpha}^{2}_{s})$}},
\newblock \bibinfo{journal}{Phys. Rev. Lett.} \bibinfo{volume}{98}
  (\bibinfo{year}{2007}) \bibinfo{pages}{022002}.
\bibitem[{Bennett et~al.(2004)}]{Bennett:2004pv}
\bibinfo{author}{G.~W. Bennett}, et~al.,
\newblock \bibinfo{title}{{Measurement of the negative muon anomalous magnetic
  moment to 0.7-ppm}},
\newblock \bibinfo{journal}{Phys. Rev. Lett.} \bibinfo{volume}{92}
  (\bibinfo{year}{2004}) \bibinfo{pages}{161802}.
\bibitem[{Aaltonen et~al.(2008)}]{:2007kv}
\bibinfo{author}{T.~Aaltonen}, et~al.,
\newblock \bibinfo{title}{{Search for $B^0_{s} \to \mu^{+} \mu^{-}$ and
  $B^0_{d} \to \mu^{+} \mu^{-}$ decays with $2fb^{-1}$ of $p \bar{p}$
  collisions}},
\newblock \bibinfo{journal}{Phys. Rev. Lett.} \bibinfo{volume}{100}
  (\bibinfo{year}{2008}) \bibinfo{pages}{101802}.
\bibitem[{Barate et~al.(2003)}]{Barate:2003sz}
\bibinfo{author}{R.~Barate}, et~al.,
\newblock \bibinfo{title}{{Search for the standard model Higgs boson at LEP}},
\newblock \bibinfo{journal}{Phys. Lett.} \bibinfo{volume}{B565}
  (\bibinfo{year}{2003}) \bibinfo{pages}{61--75}.
\bibitem[{Yao et~al.(2006)}]{Yao:2006px}
\bibinfo{author}{W.~M. Yao}, et~al.,
\newblock \bibinfo{title}{{Review of particle physics}},
\newblock \bibinfo{journal}{J. Phys.} \bibinfo{volume}{G33}
  (\bibinfo{year}{2006}) \bibinfo{pages}{1--1232}.
\bibitem[{Ahmed et~al.(2009)}]{Ahmed:2008eu}
\bibinfo{author}{Z.~Ahmed}, et~al.,
\newblock \bibinfo{title}{{Search for Weakly Interacting Massive Particles with
  the First Five-Tower Data from the Cryogenic Dark Matter Search at the Soudan
  Underground Laboratory}},
\newblock \bibinfo{journal}{Phys. Rev. Lett.} \bibinfo{volume}{102}
  (\bibinfo{year}{2009}) \bibinfo{pages}{011301}.
\bibitem[{Aprile et~al.(2010)}]{Aprile:2010um}
\bibinfo{author}{E.~Aprile}, et~al.,
\newblock \bibinfo{title}{{First Dark Matter Results from the XENON100
  Experiment}},
\newblock \bibinfo{journal}{Phys. Rev. Lett.} \bibinfo{volume}{105}
  (\bibinfo{year}{2010}) \bibinfo{pages}{131302}.
\bibitem[{Abdo et~al.(2010)}]{Abdo:2010dk}
\bibinfo{author}{A.~A. Abdo}, et~al.,
\newblock \bibinfo{title}{{Constraints on Cosmological Dark Matter Annihilation
  from the Fermi-LAT Isotropic Diffuse Gamma-Ray Measurement}},
\newblock \bibinfo{journal}{JCAP} \bibinfo{volume}{1004} (\bibinfo{year}{2010})
  \bibinfo{pages}{014}.

\end{thebibliography}

\end{document}